# General procedure for solution of contact problems under dynamic normal and tangential loading based on the known solution of normal contact problem

Valentin L Popov, Roman Pohrt, Markus Heß
Institur für Mechanik, Technische Universität Berlin, Germany
**Corresponding author**: Roman Pohrt, Technische Universität Berlin, Sek. C8-4, Straße des 17 Juni 135, 10623 Berlin, Germany.
Email: roman.pohrt@tu-berlin.de, Tel: +49 30 31421492

**Keywords**
Friction, sliding, contact, Method of Dimensionality Reduction (MDR), Mindlin

**Abstract**
In the present paper we show that the normal contact problem between two elastic bodies in the half-space approximation can always be transformed to an equivalent problem of the indentation of a profile into an elastic Winkler foundation. Once determined, the equivalent profile can be used also for tangential contact problems and arbitrary superimposed normal and tangential loading histories as well as for treating of contact problems with linearly viscoelastic bodies. In the case of axis-symmetric shapes, the equivalent profile is given by the MDR integral transformation. For all other shapes, the profile is deduced from the solution of the elastic contact normal problem, which can be obtained numerically or experimentally.

## Introduction

Contact mechanics and friction play a key role in many technological and biological systems. Due to the multiscale roughness of the contacting surfaces the treatment of associated contact problems proves to be difficult. Even the simple case of a non-adhesive, frictionless normal contact problem between two linear elastic solids with randomly rough surfaces is still a controversial scientific issue. Several analytical and numerical methods were developed to deal with the normal contact problem. Usually, numerical calculations are based on Finite Element Method [1], Boundary Element Method [2] or Green`s Function Molecular Dynamics [3], each of which has certain advantages over the other methods. A broad overview with discussion of the existing numerical and analytical methods can be found in [4].

It is self-explanatory that the presence of friction makes the contact problem more complicated. In the classical uncoupled tangential contact problem between two linear elastic spheres Cattaneo [5] and Mindlin [6] assumed a constant normal force $F_N$ and a subsequently applied, increasing tangential force $F_x$. It is well-known that this kind of loading results in the formation of a slip domain near the boundary of the contact area, while the inner domain remains in stick. However, this tangential contact problem becomes more complex for arbitrary loading scenarios since the state of stress depends not only upon the initial state of loading but also upon the complete loading history [7].

One could assume that considering the tangential contact of nominally flat rough surfaces under arbitrary varying loads will increase the difficulty of the contact problem even further. However, this is not the case, due to the Ciavarella-Jäger theorem. Jäger [8][9] and Ciavarella [10][11] independently showed that the tangential stresses in the tangential contact problem are equivalent to the difference between the actual normal stresses and those that correspond to a smaller contact area (the stick area), both multiplied with the coefficient of friction. The Ciavarella-Jäger theorem holds for all two-dimensional contact problems between solids of elastic similar materials irrespective of whether the contact area is simply connected or even spread over multiple spots. For all three-dimensional contact problems of elastic similar bodies including the classical problem of Cattaneo and Mindlin, it only applies in an approximate sense. The classical problem states that the frictional stresses in the slip domain are all directed in the direction of the applied tangential force. With the exception of the unrealistic case of $v_1 = v_2 = 0$, this assumption violates the condition that at every point in the slip domain, the slip opposes the direction of tangential stress. The reason for this is the presence of an additional deformation perpendicular to the direction of the applied force. For the classical contact of parabolic bodies, however, it could be proven that this component may be neglected [12][13]. We assume that this approximation is also valid for the generalization of the Cattaneo-Mindlin theory for arbitrary contacts including contact between bodies with randomly rough surfaces. For the latter case, a series of papers [14][15][16] investigated the frictional energy dissipation generated by varying normal and tangential forces by use of the theorem.

A further immediate consequence of the Ciavarella-Jäger theorem is the possibility of replacing the contact problem of an approximately isotropic surface shape by an equivalent axis-symmetric contact problem. Provided that the elastic normal contact problem has been solved, the equivalent profile can be deduced starting from the Galin-Sneddon integral equation [17][18]. Aleshin et al. [19] followed this way and studied the tangential contact of the equivalent axis-symmetric profile for arbitrary loading scenarios by the method of memory diagrams (MMD). In contrast to the work of Mindlin and Deresiewicz [7], the MMD replaces the complex traction distribution inside the contact area by a simple internal function containing the same memory information. Therefore the MMD is a powerful tool to calculate the hysteretic tangential force displacement curves resulting from an arbitrary loading scenario of frictional contact problems.

For the case of axis-symmetric profiles, the Method of Dimensionality Reduction (MDR) [20][21][22] is an elegant and powerful procedure for evaluating both normal and tangential contact. It starts by generating a one-dimensional profile which corresponds to the axis-symmetric shape. Following the ideas of Lee and Radok [23], the MDR can also be used for solving normal contact problems that involve linear viscoelastic media.

We thus know that for every arbitrary three-dimensional contact there is an equivalent axis-symmetric problem and that for any axis-symmetric shape there is an equivalent one-dimensional profile.

In this paper we will show a generalized rule for obtaining the equivalent one-dimensional profile which only depends on the original 3D geometry. Once the one-dimensional equivalent profile is found, the numerical procedures of MDR can be applied for both normal and tangential contact. These procedures consist of only linear equations with independent degrees of freedom. In the first section we will show how a general 1D profile is obtained from a known solution of the frictionless elastic indentation problem. In the following section, it will be displayed why and how this equivalent profile can be used in order to simulate the dynamic tangential contact. Then we will focus on how to obtain the equivalent profile for different geometries. Finally we will show some numerical examples.

**Equivalent elastic foundation and equivalent profile**

Consider a contact between an elastic indenter of arbitrary shape $z = f(x,y)$ with an elastic half space. From the contacting bodies' Youngs moduli $E_1$ and $E_2$, poisson ratios $v_1$ and $v_2$ and moduli of shear $G_1$ and $G_2$, we define the reduced moduli

$$E^* = \left( \frac{1-v_1^2}{E_1} + \frac{1-v_2^2}{E_2} \right)^{-1} \quad \text{and} \quad G^* = \left( \frac{2-v_1}{4G_1} + \frac{2-v_2}{4G_2} \right)^{-1} \tag{1}$$

During the indentation, the normal force $F_N$ is a continuous, monotonically increasing function of the indentation depth $d$. Therefore we can define unambiguously an incremental stiffness

$$k = \frac{dF_N}{dd} \tag{2}$$

which can also be expressed as a unique function of the indentation depth

$$k = k(d) . \tag{3}$$

Let us introduce formally the 'contact length', sometimes called 'Holm-radius' in literature

$$l = \frac{k}{2E^*} \tag{4}$$

The indentation depth is a unique function of the contact stiffness and thus of the contact length l :

$$d = g(l) \tag{5}$$

Note that l has the unit length and depends only on the topography and the indentation depth (also in unit length). Equation (5) thus links only geometrical quantities, independently on material properties.
Let us consider the process of indentation from its very first moment until the final indentation depth $d$, the current values of the normal force and indentation depth being given by $\tilde{F}_N$, $\tilde{d}$. During the indentation the indentation depth changes from $\tilde{d} = 0$ to $\tilde{d} = d$, the normal force changes from $\tilde{F}_N = 0$ to $\tilde{F}_N = F_N$, and the contact length from $\bar{l} = 0$ to $\bar{l} = l$. The final normal force can be written as

$$F_N = \int_0^{F_N} d\tilde{F}_N = \int_0^l \frac{d\tilde{F}_N}{d\tilde{d}} \frac{d\tilde{d}}{d\bar{l}} d\bar{l} = \int_0^l \bar{k} \frac{d\tilde{d}}{d\bar{l}} d\bar{l} = 2E^* \int_0^l \bar{l} \frac{dg(\bar{l})}{d\bar{l}} d\bar{l} \tag{6}$$

which gives after partial integration

$$F_N = 2E^* \left[ l \cdot g(l) - \int_0^l g(\bar{l}) d\bar{l} \right] = 2E^* \int_0^l \left( d - g(\bar{l}) \right) d\bar{l} \tag{7}$$

This equation can be easily interpreted as a force resulting from the indentation of the profile (5) into an elastic foundation as defined by the Method of Dimensionality Reduction [21].

Indeed, consider an elastic foundation of independent springs with equal distance $\Delta x$, each having the normal stiffness

$$\Delta k_z = E^* \Delta x \tag{8}$$

as depicted in Figure 1. The tangential stiffness of each spring is given by

$$\Delta k_x = G^* \Delta x \tag{9}$$

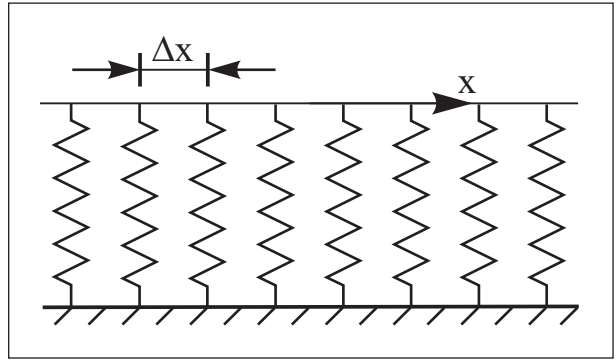

**Figure 1**. Equivalent elastic foundation

If the profile $g(x)$ is pressed into the elastic foundation defined by (8) the surface displacement in normal direction at any point *x* will be given by the difference of the indentation depth *d* and the profile shape $g(x)$:

$$u_z^{1D}(x) = d - g(x) \tag{10}$$

For contacts without adhesion, the displacement vanishes at the edge of the contact:

$$u_z^{1D}(l) = d - g(l) = 0 \tag{11}$$

The normal force in a single spring is given by

$$\Delta F_N(x) = \Delta k_z (d - g(x)) = E^*(d - g(x))\Delta x \tag{12}$$

from which the total normal force in the equilibrium state can be calculated by summation over all springs. In the limiting case $\Delta x \to 0$ the sum will be the integral

$$F_N = E^* \int_{-1}^{1} u_z^{1D}(x) dx = 2E^* \int_0^1 (d - g(x)) dx \qquad (13)$$

It can be seen easily, that the equations (11) and (13) reproduce (5) and (7). Therefore the profile $g(x)$ is the geometrical interpretation of the dependence $d = g(l)$ for the given three-dimensional profile shape.

In order to generate the equivalent profile for a given three-dimensional topography, three different procedures are at our disposal.
When the original indenting shape is an axis-symmetric profile $f(r)$ which depends only the radial coordinate $r$ and has a compact (circular) contact area, then the equivalent profile $g(x)$ is given by the MDR transformation

$$g(x) = |x| \int_0^{|x|} \frac{f'(r)}{\sqrt{x^2 - r^2}} dr \qquad (14)$$

stemming from the well-known solution of Galin-Sneddon for the normal contact problem of axis-symmetric profiles. One can either evaluate equation (14), or compose the equivalent profile using a Taylor series of $f(r)$. For more details, see [21] chapter 3.

In the case of non-axis-symmetric profiles the equivalent profile also does always exist but the transformation rule is generally not known. In some special cases an equivalent profile can be found also for complicated, non-axisymmetric surface geometries. This is the case when an analytical solution of the normal indentation is available. Consider for instance fractal rough surfaces with given Hurst Exponent $H$. It has been shown in [24] that here (with some statistical deviation stemming from the randomness) the normal Force depends on the indentation depth as

$$F(d) \propto d^{\frac{H+1}{H}} \qquad l \propto d^{\frac{1}{H}} \qquad (15)$$

We can thus derive the equivalent profile in the form of $g(x) = \text{const} \cdot x^H$.
In all other cases, the dependency between the Holm radius and the indentation depth can be obtained experimentally or numerically. The boundary elements method is suitable for the later and some examples of such simulations and their respective equivalent profiles can be found in the section 4.
For the experimental approach, the dependency can be found by indenting the original shape into a soft linear elastic counterpart such as a silicon rubber and recording both the penetration depth and the resulting normal force. The derivative of the normal force normalized by the effective Young's modulus then gives the Holm radius.

## Solution of tangential contact using equivalent profiles

As shown in [21], the tangential contact can be described in the frame of the MDR by assuming for the interaction of springs in the equivalent MDR-model the Coulomb's law of friction with the same coefficient of friction as in the original three-dimensional contact problem. That is, it is assumed that a spring sticks to the profile if the tangential force caused by the tangential displacement of the profile does not exceed the normal force acting in this spring multiplied by the coefficient of friction, and it is equal to the normal force multiplied to the coefficient of friction in the sliding region, see Figure 2.

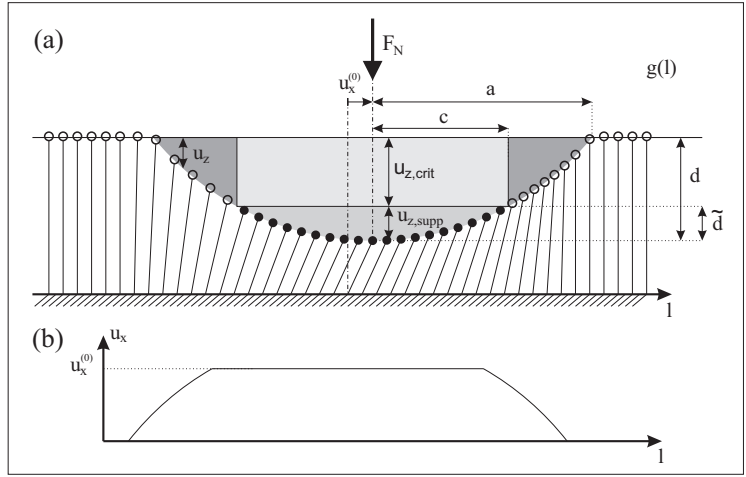

**Figure 2. (a)** MDR-model for the normal and tangential contact. The transformed shape $g(l)$ is pressed into the foundation of independent springs, shown as lines. The deflection $u_z(l)$ depends on the global indentation depth $d$ and the local value of $g(l)$. When a tangential motion is imposed, some springs stick (full circles) and some slip (open circles). **(b)** Local tangential deflection $u_x(l)$ for the above contact. Springs in the stick zone take the value of $u_x^{(0)}$. All other springs are in sliding state, and their tangential displacement is equal to $u_x(l) = u_z(l)(\mu E^* / G^*)$.

It was proven in [21] for arbitrary axis-symmetrical profiles that application of this rule reproduces the solution of Cattaneo/Mindlin and satisfies the Ciavarella/Jäger superposition principle. Below we will show that this procedure is valid also in the general case of arbitrary topographies by using the equivalent profile obtained according to equation (5).

The simplest way to show this is to go from the Ciavarella/Jäger principle, which states that the tangential stress in a tangential contact with partial sliding can be expressed as

$$\tau(x, y) = \begin{cases} \mu p_l(x, y) & \text{where sliding occurs} \\ \mu \cdot (p_l(x, y) - p_c(x, y)) & \text{where sticking occurs} \end{cases} \qquad (16)$$

where $p_l(x,y)$ is the pressure distribution in the current state which we can unambiguously characterize by the Holm radius $l$ (hence the index '$l$'). $p_c(x,y)$ is a corrective pressure distribution which is also a solution of the normal contact problem with the same geometry but a different indentation depth and thus corresponding to a different Holm radius, which we denote $c$. Integrating over the whole contact region, we get for the total tangential force

$$F_x = \mu \left( F_N(l) - F_N(c) \right) \tag{17}$$

where $F_N(l)$ is the normal contact force in the current state (corresponding to the Holm radius $l$, and $F_N(c)$ is the normal contact force corresponding to the Holm radius $c$ of the stick region.

In the equivalent MDR system, the very same principle is true. When a tangential deflection $u_x^{(0)}$ is imposed, all springs whose the tangential force is smaller than the normal force multiplied with the coefficient of friction will stick. The boundary of the stick region is given by the equality of the tangential force to the normal force times coefficient of friction: $G^* u_x^{(0)} = \mu E^* u_z(c)$. The tangential deflections outside the stick region are given by the condition $G^* u_x(x) = \mu E^* u_z(x)$. With $C_m = E^*/G^*$, the tangential force thus can be written as

$$\begin{aligned} F_x &= 2\int_0^l G^* u_x(x)dx = 2G^* \left( \int_0^c u_x(x)dx + \int_c^l u_x(x)dx \right) \\ &= 2G^* \left( \int_0^c \mu C_m u_z(c)dx + \int_c^l \mu C_m u_z(x)dx \right) \\ &= 2E^* \mu \left( \int_0^c (u_z(c) - u_z(x))dx + \int_0^l u_z(x)dx \right) \\ &= \mu \left( F_N(l) - F_N(c) \right) \end{aligned} \tag{18}$$

which coincides with the Ciavarella and Jäger result, equation (17). A more detailed derivation of this result including the Ciavarella and Jäger superposition principle is given in the supplemental material to this article.

### Examples of equivalent profiles

In the previous section we discussed how to find equivalent profiles for different original topographies. Here we generate and discuss the equivalent profiles for selected cases which are not covered by the MDR-transformation. All solutions are obtained using the boundary elements method as described in [2]. It iteratively finds a subset of discrete grid points in contact which satisfies the boundary conditions of having zero gap width inside and vanishing pressure outside the contact zone. Every subset of grid points defines a (not necessarily connected) area of contact from which one calculates the current Holm radius. In principle, one might as well record the normal force as the integral over the surface pressure and use its derivative with respect to $d$. In the following figures we show the indentation depth over the contact length (Holm-Radius) for different topographies as well as a plot of $-z(x,y)$ for $y=0$ and $x = [0 \ldots L/2]$ for comparison.

In Figure 3, we have chosen an axis-symmetric, shifted profile (see thin line in Figure 3 b) which cannot be transformed using equation (14) because the resulting contact area at low loads is ring-shaped and thus not compact. In the equivalent profile one can see that the Holm radius very quickly takes the value of flat torus radius (0.53 in the scaling). This is to be expected as the Holm radius is very much dominated by the maximum spatial spread of the contact region.

In Figure 4 we have generated a sinusoidal profile consisting of 9 peaks that first enter into contact at isolated spots and later merge into a bigger contact area. This transition is indeed visible in the equivalent profile near $l=1$. The maximum Holm radius that can be reached is given by the square comprising all peaks ($l=1.1530L$ where $L=1$ is the edge of the square, not in plot).

The topography shown in Figure 5 is randomly rough and self-affine with Hurst exponent $H=1$. As expected from equation (15), the resulting equivalent profile is approximately linear and only transitions to the saturation value of $l=L$ at large $d$.

Figure 6 shows a similar case where the roughness is applied onto a parabolic shape. There is a general semi-analytical solution available for particular cases of this scenario [25] with fractal roughness in the absence of a long-wavelength cutoff in the power spectrum. However, in the current example there is such a cutoff which makes the roughness appear nominally flat. Therefore, no analytical solution for the normal contact is available. The equivalent profile is instead obtained through direct simulation. The curve shows a transition to a $d \propto l^2$ dependency when the parabolic shape dominates the indentation behavior at high $d$. Ultimately a saturation in $l$ is reached due to the finite shape.

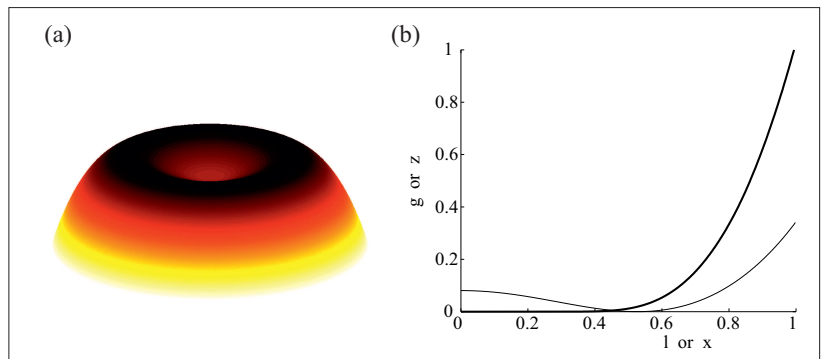

**Figure 3. (a)** Three-dimensional representation of an example of an axis-symmetric indenter shape. **(b)** Original profile as section of the body (thin line, negative sign) and equivalent MDR profile according to equation (5) (bold line).

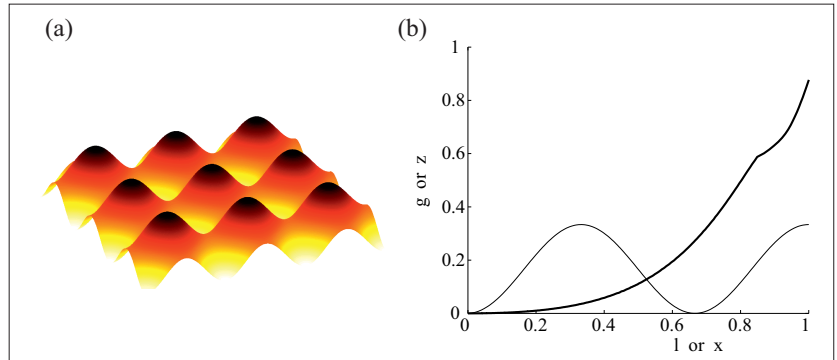

**Figure 4. (a)** Three-dimensional representation of an example of an ondulated shape. **(b)** Original profile as section of the body (thin line, negative sign) and equivalent MDR profile according to equation (5) (bold line).

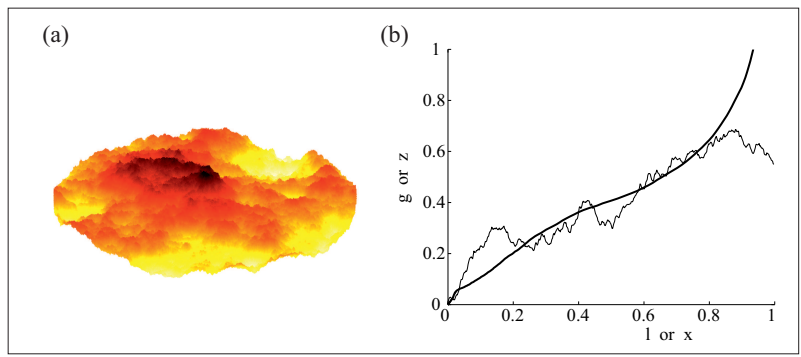

**Figure 5. (a)** Three-dimensional representation of an example of a randomly rough indenter shape. **(b)** sample section of the body (thin line, negative sign) and equivalent MDR profile according to equation (5) (bold line).

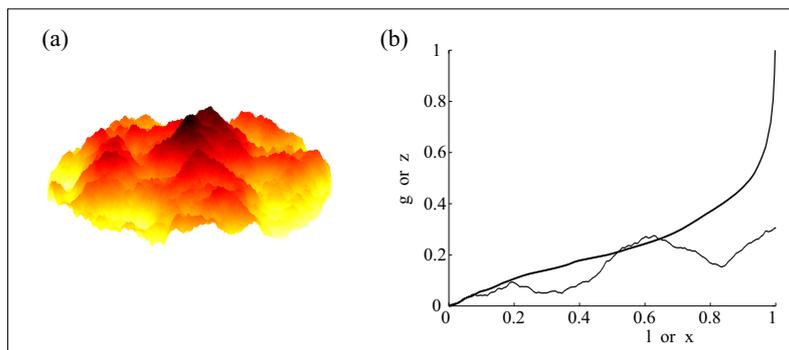

**Figure 6. (a)** Three-dimensional representation of an example of a rough parabolic indenter shape. **(b)** Original profile as sample section of the body (thin line, negative sign) and equivalent MDR profile according to equation (5) (bold line).

## Numerical sample simulation

In order to show the applicability of the proposed method, we now show an example for the tangential contact including loading history. The sample surface is the one depicted in Figure 6. All movements take place at a constant indentation depth. The resulting contact area is constant and can be seen in the red and green spots in the right column of Figure 7.

This indented surface is subjected to an oscillating tangential movement with growing amplitude. We have simulated this case using Boundary Elements Methods with Cattaneo/Mindlin principle. For every time step, we have recorded the distribution of stick and slip area and the resulting tangential force, which is shown in Figure 8.

We then used the equivalent profile in order to simulate the same tangential movements according to the rules of the MDR (see above). In Figure 7, the evolution of the spring deflections are easily interpreted. Because Coulomb friction is assumed, all deflections cannot exceed $|u_x| \leq u_z \mu C_m$. During tangential motion, the curve is simply shifted upwards or downwards, restricted by this boundary. The tangential force is obtained by evaluating the grey area. Points A and B show states shortly before and after the direction of motion is changed. In A, most of the contact zone slips. After the direction is changed (B), most points deliberately follow the external movement (they stick) with the exception of very lightly loaded points in the contact zone boundary. Please note that the same can be observed in the onedimensional model. Only a small fraction of the springs is quickly limited by the $|u_x| \leq u_z \mu C_m$ condition (red circle). In state C, the curve of $u_x$ lowers again and conforms to $-u_z \mu C_m$ but still has the shape of the upper bound in all springs that are still in sticking state (left).

Figure 8 also shows the force-displacement-dependency of the MDR calculation. Both curves are hardly distinguishable. However, the MDR procedure is dramatically simpler and requires only negligible computing time. The tangential force $F_t$ is normalized by the maximum value it can attain (COF times normal force). The tangential bulk displacement $u_{x,0}$ is normalized by it maximum value prior to macroscopic slip $u_{x,max}$. For isotropic, elastic contact, this is given by $u_{x,max} = d\mu C_m$ (see [26] for details).

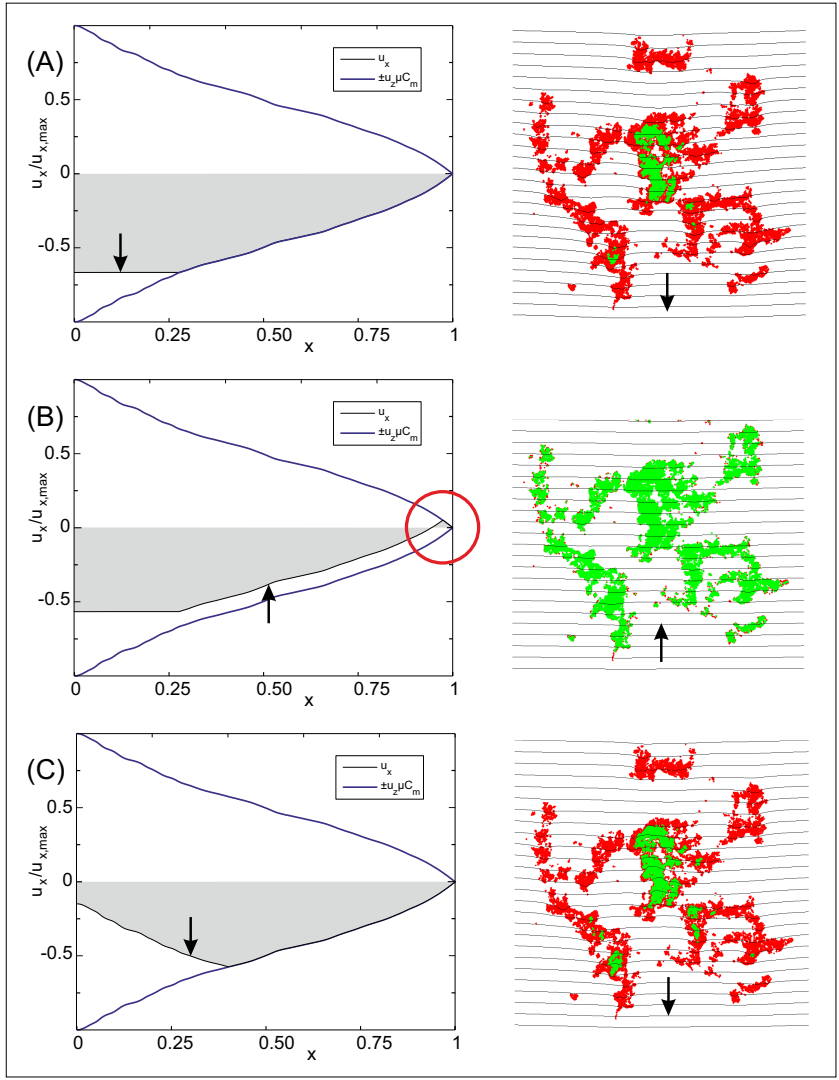

**Figure 7. (right column)** plot of the contact region of the surface shown in Figure 6. Green points are in sticking state, red points slide. We also show the tangential deflection of the outer surface by the horizontal lines. The states A-C correspond to the points marked in **Figure 8**. **(left column)** plot of the tangential deflection of the independent springs in the one-dimensional model.

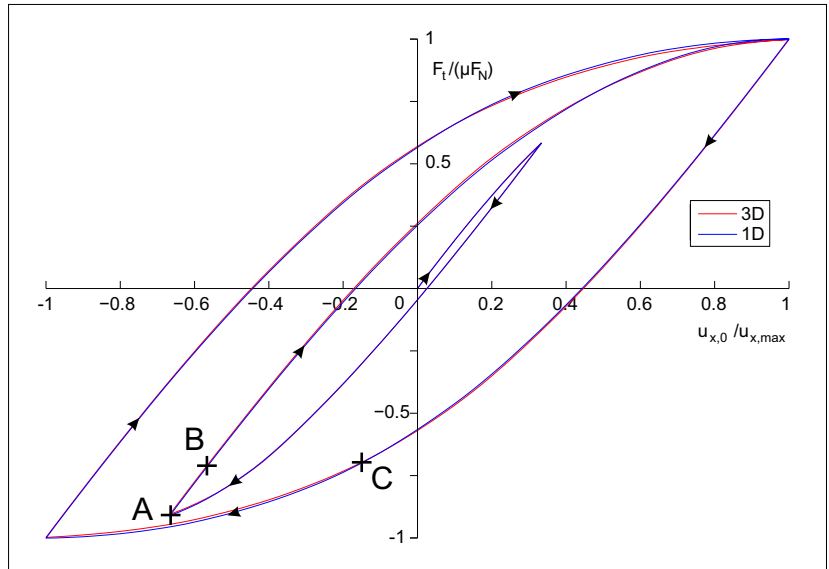

**Figure 8.** Tangential Force-displacement-curve for the tangential contact of the surface shown in Figure 6. The dependencies are generated using 3D Boundary Elements Method (red curve) and the 1D Method of Dimensionality Reduction based on the equivalent profile from Figure 6 (b).

## Discussion

The Method of Dimensionality Reduction is an easy and effective method for treating various classes of contact problems by mapping them to the contact of a modified profile with a linear elastic (or viscoelastic) foundation. Often it is erroneously believed that the MDR is only applicable to axis-symmetric profiles. This is not correct. In the present paper we have shown that the "equivalent one-dimensional profile" does exist for absolutely every arbitrary surface shape. We have shown that this profile can be found from the known solution of the normal contact problem. The one-dimensional profile is obtained directly by plotting the indentation depth over the contact length, when the dependency is known from analytical, numerical or experimental findings. We presented a numerical example how once obtained, the profile can be used also for the simulation of the tangential contact problem with a constant coefficient of friction under arbitrary loading history. Another application is the normal contact problem of any profile with an elastomer having arbitrary linear rheology. Also in this case, the MDR provides a powerful method which is easy to implement.

All presented results are correct within the usual assumptions of linear contact mechanics (half-space approximation, geometrical linearity, Mindlin/Cattaneo-approximation and assumption of uncoupling of normal and tangential problems as well as neglecting the orthogonal slip).

It would be interesting to further investigate the applicability for wear prediction, currently shown only for axis-symmetric three-dimensional shapes [27][28].

# General procedure for solution of contact problems under dynamic normal and tangential loading based on the known solution of normal contact problem

Valentin L Popov, Roman Pohrt, Markus Heß
Institut für Mechanik, Technische Universität Berlin, Germany

**Corresponding author**: Roman Pohrt, Technische Universität Berlin, Sek. C8-4, Straße des 17 Juni 135, 10623 Berlin, Germany.
Email: roman.pohrt@tu-berlin.de, Tel: +49 30 31421492

## Supplemental proof

In this Appendix we will derive the Ciavarella-Jäger superposition principle for tangential contact of arbitrary topographies and show that the usual MDR procedure can be applied to the contact of arbitrary three-dimensional topographies as soon as the equivalent MDR profile has been found.

There is a very close analogy between normal and tangential contact problems. If some flat region of the surface of an elastic half-space (of arbitrary form, not necessarily compact) is indented by the indentation depth $d$, then the normal force is proportional to the indentation depth and according to the definition, the Holm radius $l$ is given by

$$F_N = 2E^* l d \tag{1}$$

The corresponding pressure distribution $p(x, y)$ is also proportional to the indentation depth and the effective modulus of elasticity and can be written as

$$p(x, y) = E^* d \cdot \pi(x, y) \tag{2}$$

where $\pi(x, y)$ is the "reduced pressure" having the unit of $m^{-1}$ which depends only of the geometrical configuration of the considered region. The analogy to the tangential contact problem can be formulated as follows: If the region of the same form is rigidly displaced in tangential direction by $u_x^{(0)}$, then the stress distribution in the contact area will be given in good approximation by the equation

$$\tau(x, y) = G^* u_x^{(0)} \cdot \pi(x, y) \tag{3}$$

That is, it has the same form as in the case of vertical indentation, only with another elastic constant. From equation (3) it follows immediately that the tangential force is equal to

$$F_x = 2G^* l u_x^{(0)} \tag{4}$$

and that the ratio of normal and tangential stiffness is equal to $E^*/G^*$ independently on the form of the considered surface region. For the case of round contact regions, this ratio was first found by Mindlin [1], we therefore call it "the Mindlin ratio". The independence of the ratio of contact stiffnesses on the configuration of the region was proven for arbitrary *two dimensional* contacts by Ciavarella [2]. Even though in two dimensions, the indentation depth cannot be determined uniquely, the contact stiffnesses are well-defined properties. In three dimensions, the same ratio was found analytically [3] and numerically [4],[5] for randomly rough surfaces.

In the following, we go from the approximate validity of the Equations (2) and (3). As a preliminary step, consider formally a simultaneous indentation of an arbitrary profile $z = f(x,y)$ in the normal and tangential directions without slip. Let us characterize both normal and tangential displacements of the indenter as functions of the current Holm radius $l$:

$$d = g(l) \qquad u_x^{(0)} = h(l) \tag{5}$$

The normal and tangential forces and the corresponding stress components can be written

$$F_N = 2E^* \int_0^l \bar{l} \frac{dg(\bar{l})}{d\bar{l}} d\tilde{l} \qquad p(x,y) = E^* \int_r^l \pi(\bar{l};x,y) \frac{dg(\bar{l})}{d\bar{l}} d\bar{l} \tag{6}$$

and

$$F_x = 2G^* \int_0^l \bar{l} \frac{dh(\bar{l})}{d\bar{l}} d\tilde{l} \qquad \tau(x,y) = G^* \int_r^l \pi(\bar{l};x,y) \frac{dh(\bar{l})}{d\bar{l}} d\bar{l} \tag{7}$$

Here $\pi(\bar{l};x,y)$ is the current reduced stress distribution according to the definition (2) in the state corresponding to the contact length $\bar{l}$, and $r$ is the value of the contact length for the indentation at which the point $(x,y)$ is placed on the boundary of the contact region. Note that here one could also start the integration from 0, as $\pi(s,x,y) = 0$ for every $s < r(x,y)$.

Let us now consider the following two-step indentation. The indenter first is indented normally until the Holm radius $c$ is achieved. After that, it is indented up to the Holm radius $l$ simultaneously in the normal and tangential directions so that

$$dh = \lambda \cdot dg \tag{8}$$

The normal force and the pressure distribution in the normal direction will be still given by equation (7) while for the tangential force we get

$$F_x = 2G^* \int_c^1 \bar{l} \frac{dh(\bar{l})}{d\bar{l}} d\tilde{l} \qquad (9)$$

and for the tangential stress component

$$\tau(x,y) = \begin{cases} G^* \int_c^1 \pi(\bar{l};x,y) \frac{dh(\bar{l})}{d\bar{l}} d\bar{l}, & \text{for } r < c \\ G^* \int_r^1 \pi(\bar{l};x,y) \frac{dh(\bar{l})}{d\bar{l}} d\bar{l}, & \text{for } c < r < 1 \end{cases} \qquad (10)$$

For any point lying outside the boundary of the contact region corresponding to the Holm radius $c$ and inside the region corresponding to the Holm radius $l$, the distributions of the tangential and the normal stress have the same form:

$$\tau(x,y) = \lambda \frac{G^*}{E^*} p(x,y) \qquad (11)$$

If we choose

$$\lambda \frac{G^*}{E^*} = \mu \qquad (12)$$

then the described contact will have the following properties:

$$\begin{aligned} u_x(x,y) &= u_x^{(0)} = \text{const}, && \text{for } 0 \leq r(x,y) \leq c \\ \tau(x,y) &= \mu p(x,y), && \text{for } c < r(x,y) \leq 1 \end{aligned} \qquad (13)$$

These equations correspond exactly to the stick and slip conditions in a tangential contact with the coefficient of friction $\mu$. The distribution of the tangential stress takes the form

$$\tau(x,y) = \begin{cases} \mu(p_1(x,y) - p_c(x,y)), & \text{for } 0 \leq r(x,y) \leq c \\ \mu p_1(x,y), & \text{for } c < r(x,y) \leq 1 \end{cases} \qquad (14)$$

where $p_l(x,y)$ and $p_c(x,y)$ are normal stress distributions corresponding to the Holm radii l and c respectively. The tangential displacement of the contact can be obtained by integrating (8) and substituting (12)

$$u_x^{(0)} = \mu \frac{E^*}{G^*}\left(g(a) - g(c)\right) \tag{15}$$

These equations coincide with equations of the Method of Dimensionality Reduction initially derived for axis-symmetric contacts [6]. The above derivation shows that they are valid for contacts of arbitrary surface topography.